\documentclass[runningheads]{llncs}
\pdfoutput=1
\usepackage{amsmath}
\usepackage[T1]{fontenc}
\usepackage{graphicx}
\usepackage[%
    font={small}
]{caption}
\usepackage[list=true]{subcaption}
\usepackage{xcolor} % For text colouring
\usepackage{hyperref} % For hyperlinks
\begin{document}
\title{Diffusion-based Adversarial Purification for Intrusion Detection}
\author{Mohamed Amine Merzouk\inst{1,2}\and
Erwan Beurier\inst{1,2}\and
Reda Yaich\inst{2}\and
Nora Boulahia-Cuppens\inst{1}\and
Frédéric Cuppens\inst{1}}
\authorrunning{Merzouk \textit{et al.}}
\institute{Polytechnique Montréal, Montréal, Canada\\
\email{\{mohamed-amine.merzouk,erwan.beurier,\\
nora.boulahia-cuppens,frederic.cuppens\}@polymtl.ca}
\and
IRT SystemX, Palaiseau, France\\
\email{\{mohamed-amine.merzouk,erwan.beurier,\\
reda.yaich\}@irt-systemx.fr}}
\maketitle              % typeset the header of the contribution

\begin{abstract}
The escalating sophistication of cyberattacks has encouraged the integration of machine learning techniques in intrusion detection systems, but the rise of adversarial examples presents a significant challenge.
These crafted perturbations mislead ML models, enabling attackers to evade detection or trigger false alerts.
In response, adversarial purification has emerged as a compelling solution, particularly with diffusion models showing promising results.
However, their purification potential remains unexplored in the context of intrusion detection.
This paper demonstrates the effectiveness of diffusion models in purifying adversarial examples in network intrusion detection.
Through a comprehensive analysis of the diffusion parameters, we identify optimal configurations maximizing adversarial robustness with minimal impact on regular performance.
Importantly, this study reveals insights into the relationship between diffusion noise and diffusion steps, representing a novel contribution to the field.
Our experiments are carried out on two datasets and against five adversarial attacks.
The implementation code is publicly available. 
\keywords{Adversarial defense \and Adversarial purification \and Adversarial examples \and Diffusion models \and Intrusion detection.}
\end{abstract}

%% \backgroundsetup{
%%       contents={\textbf{CONFIDENTIAL: DO NOT DISTRIBUTE OR CITE.}},
%%       color=gray,
%%       scale=2.5
%%     }%

\section{Introduction}

Intrusion detection stands out as one of the most formidable challenges in cybersecurity, especially with the increasing sophistication of cyberattacks.
Unfortunately, traditional signature-based approaches reach their limit against previously unknown threats, also called zero-days.
As such, the integration of Machine Learning (ML) techniques has emerged as a promising avenue for enhancing the detection capabilities of intrusion detection systems.

However, the advent of adversarial examples~\cite{szegedy_intriguing_2014,goodfellow_explaining_2015} poses a severe obstacle to the reliability of ML, specifically in critical tasks such as intrusion detection.
They are generated from regular data instances by adding a meticulously crafted perturbation that misleads an ML model.
Applied to network data, they enable cyber attackers to either evade ML-based intrusion detection systems or flood the network with false alerts.

In response to the escalating threat of adversarial attacks, research efforts have been directed toward designing effective countermeasures.
Among the various defensive approaches, adversarial purification has emerged as a compelling solution to remove the adversarial perturbation from data before processing it.
This defense is particularly interesting for intrusion detection, as it can be integrated upstream of the model without retraining.
Recent work~\cite{nie_diffusion_2022} has demonstrated promising purification performance using diffusion models~\cite{sohl-dickstein_deep_2015,ho_denoising_2020}.

Diffusion models are generative models inspired by the dynamics of diffusion processes in physics~\cite{sohl-dickstein_deep_2015}.
They consist of a forward process that gradually adds noise to initial data and a backward process that reconstructs that data using a deep neural network.
Because diffusion models are trained using examples drawn from the original data distribution, the reconstructed data is expected to adhere to the same distribution, even when the initial data is an adversarial example.
Thus, they can be used for adversarial purification: removing the perturbation from adversarial examples to classify them correctly.
%%Furthermore, the unsupervised nature of diffusion models enables an attack-agnostic purification.
Furthermore, since they are not explicitly trained on adversarial examples, their purification performance is not limited to a specific adversarial attack. 

Despite the recent attention given to adversarial purification with diffusion models, there remains a notable gap in understanding their potential in the context of intrusion detection.
This paper fills that gap by studying the purification effects of diffusion models and demonstrating their effectiveness on network intrusion detection.
By conducting a comprehensive analysis of the diffusion parameters, we identify optimal configurations maximizing the adversarial robustness with limited impact on the regular performance.
We show the effectiveness of our method on two network datasets (UNSW-NB15~\cite{moustafa_nsw_nb15_2015} and NSL-KDD~\cite{tavallaee_nsl_kdd_2009}) and against state-of-the-art adversarial attacks.
Moreover, to our knowledge, this is the first investigation of the relationship between the number of diffusion steps and the optimal amount of diffusion noise for adversarial purification.
Our findings demonstrate that the optimal amount of diffusion noise depends not on the number of diffusion steps but rather on the amount of adversarial perturbation.

Section~\ref{section: related work} introduces the background and related works in cybersecurity.
In Section~\ref{section: methodology}, we describe the experiment methodology.
We report the results in Section~\ref{section: results}, followed by a discussion in Section~\ref{section: discussion} and a conclusion in Section~\ref{section: conclusion}. 

\section{Background and Related Work}
\label{section: related work}

This section introduces the major approaches to adversarial defense, provides some background on diffusion models, and reviews the literature on diffusion models in intrusion detection systems and adversarial purification. 

\subsubsection{Adversarial Defenses.}

%% Adversarial examples are problematic in all applications of neural networks, and in particular, critical fields like cybersecurity, health, or autonomous vehicles; several techniques exist to handle them.
The following presents three dominant approaches to defend against adversarial examples~\cite{yuan_adversarial_2019}.

\textit{Adversarial training} consists of training the model with adversarial examples in addition to the training data~\cite{goodfellow_explaining_2015,madry_towards_2018}. 
Despite its effectiveness, this technique has several drawbacks: (\textit{i}) it requires the retraining of the model and significantly lengthens the training duration, and (\textit{ii}) it protects only against adversarial examples generated with the methods it was trained on.

\textit{Adversarial detection} consists of a separate classifier deployed upstream of the ML model that detects and discards adversarial examples before they are fed to the model~\cite{aldahdooh_adversarial_2022}.
It is plug-and-play and does not require retraining.
Several supervised and unsupervised techniques exist; interested readers may refer to~\cite{aldahdooh_adversarial_2022} for a detailed review.
Unfortunately, adversarial detection also depends on the adversarial attacks on which it was trained.
Moreover, since it is a classifier, it can be fooled to some extent~\cite{carlini_adversarial_2017}. 

% Other techniques like gradient masking exist too (but can be circumvented~\cite{athalye_obfuscated_2018})
\textit{Adversarial purification} consists of a separate model deployed upstream of the ML model that removes the perturbation from adversarial examples before they are fed to the model~\cite{song_pixeldefend_2017,srinivasan_robustifying_2021}.
This approach is also plug-and-play, and the purification models are typically trained independently of the ML model.
Adversarial purification does not require retraining, and it is, in many cases, independent of the adversarial attacks.
This paper focuses on the adversarial purification approach using diffusion models~\cite{nie_diffusion_2022}.

\subsubsection{Diffusion Models.} Diffusion models are a class of generative models that leverage the diffusion processes used in physics to learn complex data distributions and samples from them~\cite{sohl-dickstein_deep_2015,ho_denoising_2020}.
Instead of modeling the distribution like VAEs~\cite{kingma_auto-encoding_2022} and GANs~\cite{goodfellow_generative_2014}, they model the process of transforming a simple distribution (e.g., Gaussian noise) into the target distribution through a sequence of steps.
They consist of two Markov processes: a \emph{forward} and a \emph{reverse} process.

In the \emph{forward process}, each step involves adding Gaussian noise to the data over $T$ steps until no structure remains; it corresponds to a smooth transition from the complex data distribution to a Gaussian distribution (latent space).
The amount of noise added at each step depends on the variance schedule $\beta$.
In the \emph{reverse process}, each step reverts the corresponding forward step—removes the diffusion noise—to reconstruct the original data distribution.
Since the reverse process is mathematically intractable, it is approximated with deep neural networks.
In this work, we use discrete diffusion steps, but the continuous case can also apply; it requires solving stochastic differential equations~\cite{song_score-based_2021}.

%are latent variable models with both a \emph{forward process} and a \emph{reverse process}. In this section, we give a short introduction to the discrete timesteps case. We refer the interested reader to~\cite{sohl-dickstein_deep_2015} or~\cite{ho_denoising_2020} for a more detailed description, or to~\cite{song_score-based_2021} for the continuous timesteps case. 
In the following, we provide the intuition into the theory behind diffusion models.
We consider a dataset $x_0$ with unknown distribution $x_0 \sim q\left(x_0\right)$. For a given number $T$ of steps, we consider the Markov chain $\left(x_t\right)_{t \leq T}$ with transitions 

\begin{equation}
    q\left(x_{t+1} | x_t \right) = \mathcal{N}\left( x_{t+1} ; \sqrt{1 - \beta_{t+1}} x_{t} , \beta_{t+1} I_n\right),
    \label{eq:transition_one_step_forward}
\end{equation}

\noindent
that is, we gradually add Gaussian noise with a given variance $\left(\beta_t\right)_{t \leq T}$ to the data.
If we define $\bar{\alpha_t} = \prod_{i=1}^{i=t}{(1 - \beta_{i})}$, then the cumulative noise addition from the clean data to step $t$ is written:

\begin{equation}
    q\left(x_{t} | x_0 \right) \simeq \mathcal{N}\left( x_{t} ; \sqrt{\bar{\alpha_{t}}} x_{0}, \left(1 - \bar{\alpha}_t\right) I_n \right).
    \label{eq:transition_cumulative_forward}
\end{equation}

Equation \ref{eq:transition_cumulative_forward} describes the \emph{forward process} of diffusion models.
Note that to ensure the diversity of generated data the variance schedule should guarantee that the data resembles a Gaussian distribution at the end of the forward process:

\begin{equation}
    q\left(x_{T} | x_0 \right) \simeq \mathcal{N}\left( x_{T} ; 0, I_n\right).
\end{equation}

The \emph{reverse process} consists of generating examples from the original data distribution using the reverse Markov chain; it starts from a Gaussian distribution 
%%\begin{equation}
%%    p\left(x_{T} \right) = \mathcal{N}\left( x_{T} ; 0, I_n\right)
%%\end{equation}
\noindent with transitions

\begin{equation}
    p\left(x_t | x_{t+1} \right) =  \mathcal{N}\left( x_{t+1} ; \mu_{\theta}\left( x_t, t \right) , \Sigma_{\theta}\left( x_t, t \right) \right),
\end{equation}

\noindent
where $\theta$ represents the parameters of the deep neural network used to estimate the diffusion noise to be removed.

%% The idea behind diffusion models can be described as follows~\cite{song_score-based_2021}. The distribution $q\left(x_0\right)$ of the data $x_0$ is usually very complicated. The forward process converts $q\left(x_0\right)$ to another distribution $q\left(x_{T} | x_0 \right)$ by gradually adding noise. The resulting distribution is easier to manipulate or sample. The reverse process converts the sampled data back to the data distribution, again by gradually denoising the data. The resulting sample belongs to the original distribution, hence the success of diffusion models as generative models~\cite{ho_denoising_2020,dhariwal_diffusion_2021}.

%% The reverse process is mathematically intractable. However, it can be approximated with deep neural networks. For this purpose, our implementation follows~\cite[Algorithms 1]{ho_denoising_2020}, which remains relevant to our network traffic data.

\subsubsection{Adversarial Purification with Diffusion Models.}
Adversarial purification is the process of removing the perturbation from adversarial examples to classify them correctly.
This process can be seen as a generative task and approached with diffusion models.
The gradual addition of Gaussian noise in the forward step submerses the adversarial perturbation, but the data becomes too noisy to be correctly classified.
Thus, the reverse step reconstructs the data in the original distribution without adversarial perturbations.

Furthermore, the forward process does not need to complete $T$ steps and reach a Gaussian distribution.
There should be enough added noise to submerse the adversarial perturbation, but not too much, as it damages the data structure and decreases the accuracy. 
The forward process should stop at the optimal diffusion step $t^*$, where the diffusion noise suffices to remove the adversarial perturbation while preserving the structure for the classification~\cite{nie_diffusion_2022}.

After this concept was introduced in~\cite{nie_diffusion_2022}, later work leveraged guided diffusion models for adversarial purification~\cite{wang_guided_2022,wu_guided_2022}.
Authors in~\cite{lin_robust_2024} train a robust guidance with an adversarial loss and apply it to the reverse process. 
Diffusion models are also used to purify backdoors in poisoned models~\cite{shi_blackbox_2023}.

%% Develop the research gap
%% Note that most of the literature focuses on 2D or 3D images, and no reference was found on network data. 

%%Gap in the literature
%% Use on network intrusion detection

\subsubsection{Diffusion Models in Intrusion Detection.}
Intrusion detection systems benefit greatly from the automation provided by ML, including deep learning~\cite{liu_machine_2019,sohn_deepbelief_2021,he_adversarial_2023}. 
%% However, as diffusion models are pretty new, very few references to their use in intrusion detection are to be found. 
However, network intrusion datasets are often imbalanced; benign traffic outweighs malicious traffic.
Due to their generative capabilities, diffusion models are successfully applied in data augmentation for balancing network datasets~\cite{zhang_didids_2023,tang_diffusion_2023,han_mmid_bench_2024}. 
The diffusion model can also detect intrusion by learning the distribution of benign traffic.
The difference between the original and reconstructed data is then used to detect malicious traffic~\cite{wang_intrusion_2023,yang_ddmt_2023}.

However, to our knowledge, no prior research has investigated the adversarial purification potential of diffusion models in the context of intrusion detection.
This paper represents the foremost initiative to address diffusion-based adversarial purification in intrusion detection.

\section{Methodology}
\label{section: methodology}
%Here, we describe the methodology followed in our experiments.
%First, we describe the intrusion detection models and datasets, then the adversarial attacks, and the diffusion models.
% In this section, we briefly describe our methodology.

%\subsection{Our approach}

\begin{figure}[t]
    \centering
    \includegraphics[width=\textwidth]{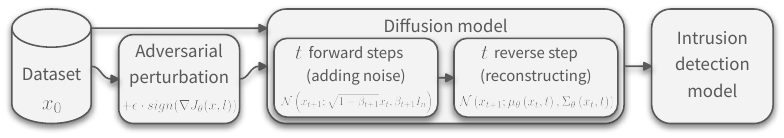}
    %\caption{Caption}
    \caption{Methodology scheme: dataset instances $x_0$ undergo adversarial perturbation, the diffusion model's purification, and then the intrusion detection classification. } 
    \label{fig:methodology}
\end{figure}

As shown in Figure~\ref{fig:methodology}, our methodology consists of a diffusion model deployed upstream of the intrusion detection model.
The diffusion model serves as a "filter" that removes adversarial perturbations.
It adds noise to the data for several steps $t$ and reconstructs it through the reverse diffusion process involving the diffusion neural network.
The data, whether original or adversarial, are first purified by the diffusion model and then fed to the intrusion detection model to determine if they are benign or malicious.

%%The intrusion detection model is taking as input network data and returning whether it is an attack or not. The adversarial attacks target the IDS. The diffusion model will add noise data, and purify it before feeding it to the classifier. The hyperparameters of the IDS are fixed along the experimentations, while we vary those of the diffusion model. 

\paragraph{The intrusion detection model} is a fully connected neural network with fixed hyperparameters throughout the experiments.
It comprises a fully connected neural network with 5 hidden layers of 256, 512, 1024, 512, and 256 Rectified Linear Units (ReLU).
It is trained for $10,000$ epochs, and the parameters are optimized using Adam with a learning rate of $10^{-5}$. 

 %In short, for each training epoch, for each vector, we randomly choose a number of noise steps $t \in [\![1, T]\!]$. Then, we add a Gaussian noise with variance $1 - \bar{\alpha_t}$. This corresponds to the sum of $t$ transitions of the forward process as described in Equation~\ref{eq:transition_cumulative_forward}. We then train our diffusion model to reconstruct the clean vector from the noisy one. 

\paragraph{The diffusion models} are trained according to~\cite[Algorithm 1]{ho_denoising_2020}.
We consider $T$ discrete diffusion steps.
Instead of having a separate neural network for each timestep~\cite{sohl-dickstein_deep_2015}, we encode the timesteps as sinusoidal embeddings and add them to each layer of the diffusion neural network~\cite{ho_denoising_2020}.

The purification results are recorded across the diffusion steps.
For each $t \in [1, T]$, we apply $t$ forward diffusion steps to the data instance $x_{0}$ to get $x_{t}$; then we apply $t$ reverse diffusion steps to $x_{t}$ to get $\hat{x}_{0}$, the reconstruction of $x_{0}$.
%In practice, we add a Gaussian noise with variance $1 - \bar{\alpha_t}$.
%This corresponds to the sum of $t$ transitions of the forward process as described in Equation~\ref{eq:transition_cumulative_forward}.

\paragraph{The variance schedule} $\beta$ is linearly distributed ($T$ evenly spaced values between $\beta_{1}$ and $\beta_{T}$).
Across different experiments, the number of diffusion steps $T$ is $100$ or $1000$, while $\beta_{1}$ varies between $10^{-5}$ and $10^{-4}$, and $\beta_{T}$ varies between $10^{-4}$ and $10^{-1}$.

\paragraph{The diffusion neural networks} are fully connected neural networks with 10 hidden layers; each hidden layer typically consists of 960 ReLU units, except for the experiments that compare neural network architectures.
The diffusion neural networks are trained for $200,000$ epochs, where each epoch consists of predicting a random step of the reverse process for each dataset instance.
The loss function is a Mean Squared Error (MSE), and the parameters are optimized using AdamW with a learning rate of $10^{-4}$.

%Once trained, we evaluate the diffusion models on two aspects: (\textit{i}) the reconstruction performance using the reconstruction loss (the MSE between the original data and the reconstructed data), and (\textit{ii}) the adversarial purification performance with the accuracy of the intrusion detection model on reconstructed data.
%The evaluation is carried out on the training set, the testing set, and the adversarial examples generated from the testing set.

\paragraph{The metrics} recorded during the experiments evaluate two aspects of diffusion models: 
(\textit{i}) the reconstruction performance by recording the reconstruction loss (MSE between the original data and the reconstructed data) for a diffusion step $t$, and (\textit{ii}) the adversarial purification performance by feeding the reconstructed data to the intrusion detection model and recording its accuracy.

\paragraph{The optimal diffusion step} $t^{*} \in [1, T]$ is the step that maximizes the intrusion detection accuracy on adversarial examples~\cite{nie_diffusion_2022}.
It should be large enough to dilute the adversarial perturbation with diffusion noise.
However, the larger it is, the more data structure it dilutes, which decreases the test accuracy.

All experiments are carried out on two prominent network datasets:
NSL-KDD~\cite{tavallaee_nsl_kdd_2009}, which is old but still widely used for benchmarking and comparison, and UNSW-NB15~\cite{moustafa_nsw_nb15_2015} which is more recent and representative of modern network traffic~\cite{ring_survey_2019}. 
For further details on the implementation of our experiments, we make our code publicly available~\footnote{\url{https://github.com/mamerzouk/adversarial-purification}}. 
%% , both described in Table~\ref{table:datasets}.

%\begin{enumerate}
%    \item{NSL-KDD~\cite{tavallaee_nsl_kdd_2009}:} a balanced version of the KDD Cup 99 data set without the redundant records~\cite{kdd_cup99}.
%    \item{UNSW-NB15~\cite{moustafa_nsw_nb15_2015}} a more recent dataset generated in a simulated environment.
%\end{enumerate}

%See Table~\ref{table:datasets} for a description of both datasets. 

%\subsubsection{NSL-KDD}
%Network Security Laboratory 
%\cite{tavallaee_nsl_kdd_2009}
%\subsubsection{UNSW-NB15}

%The UNSW-NB15 dataset~\cite{moustafa_nsw_nb15_2015} was generated in a simulated environment. The records have 49 features and cover 9 types of attacks as well as benign traffic. The training set contains 175,341 records, a testing set of 82,332 records.

%% \subsection{Adversarial Attacks}
%% \subsubsection{Fast Gradient Sign Method}
%% \subsubsection{DeepFool}
%% Il n'y a pas grand chose à dire, ça ne vaut pas toute une sousection, je la met avec les datasets
%% \subsection{Intrusion Detection Models}
%% The intrusion detection models are fully connected neural networks with 5 hidden layers of 256, 512, 1024, 512, and 256 Rectified Linear Units (ReLU).
%% They are trained for $10,000$ epochs and the parameters are optimized using Adam with a learning rate of $10^{-5}$.

%\subsection{Diffusion Models}
%\label{h3:diffusion_models}

\section{Results}
\label{section: results}

\begin{figure}[t]
\centering
\begin{subfigure}{.5\textwidth}
  \centering
  \includegraphics[width=\linewidth]{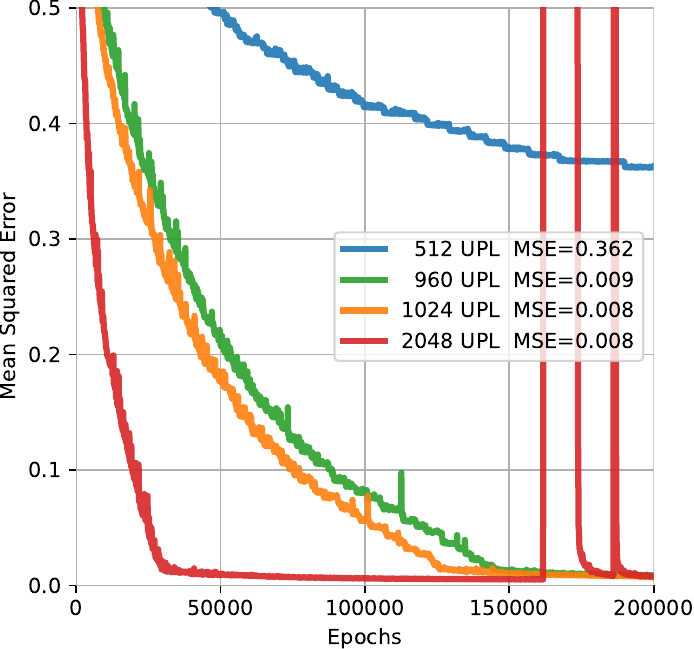}
  \caption{UNSW-NB15}
  \label{fig:loss_epochs_h_UNSW-NB15}
\end{subfigure}%
\begin{subfigure}{.5\textwidth}
  \centering
  \includegraphics[width=\linewidth]{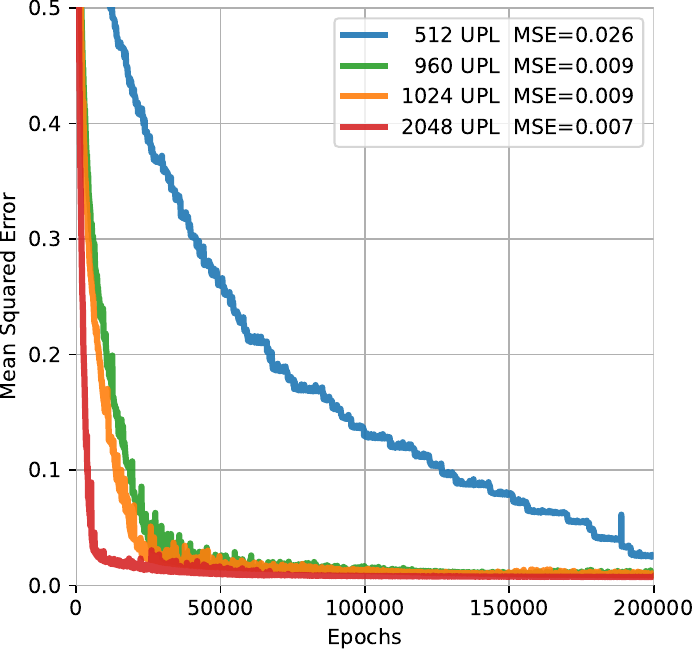}
  \caption{NSL-KDD}
  \label{fig:loss_epochs_h_NSL-KDD}
\end{subfigure}
\caption{Reconstruction loss over training epochs for different neural network sizes}
\label{fig:loss_epochs_h}
\end{figure}

In this section, we present the results of our experiments.
We first analyze the reconstruction loss throughout the training of the diffusion models to identify optimal hyperparameters.
Then, we study the reconstruction loss and the accuracy of the intrusion detection model on reconstructed data.
We compare the robustness of the intrusion detection model with respect to the level of diffusion applied to the data.
We aim to find the optimal diffusion step $t^{*}$ that removes the adversarial perturbation and increases the robustness while minimizing the repercussions on non-adversarial data.

Furthermore, we analyze the impact of several diffusion parameters on the purification performance: the number of steps $T$, the initial variance $\beta_{1}$, and the final variance $\beta_{T}$.
Finally, we study the impact of the adversarial perturbation amplitude $\epsilon$ on the optimal diffusion step $t^{*}$ and compare our purification model against five state-of-the-art adversarial attacks.
The figures present the results on both UNSW-NB15 and NSL-KDD; the values are the mean and standard deviation over 10 randomly initialized runs.

Unless otherwise noted, the diffusion models in these experiments use the standard diffusion parameters proposed in the previous work~\cite{ho_denoising_2020}: $\beta_{1}=10^{-4}$, $\beta_{T}=0.02$, and $T=1000$.
%% The diffusion neural networks are fully connected neural networks with 10 hidden layers of 960 ReLU units trained for $200,000$ epochs.

% \subsection{Reconstruction of Network Data}

%% Training cuvres for neural network sizes

%% Loss over T
\begin{figure}[t]
\centering
\begin{subfigure}{.5\textwidth}
  \centering
  \includegraphics[width=\linewidth]{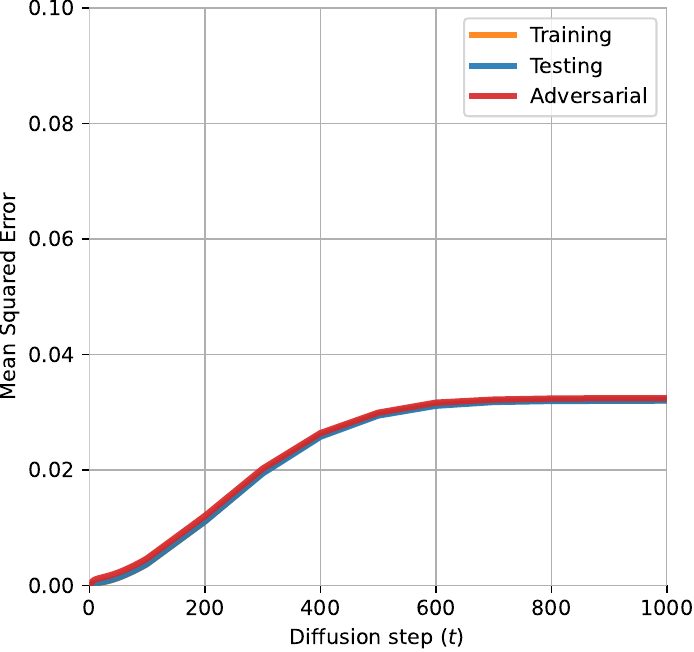}
  \caption{UNSW-NB15}
  \label{fig:loss_t_UNSW-NB15}
\end{subfigure}%
\begin{subfigure}{.5\textwidth}
  \centering
  \includegraphics[width=\linewidth]{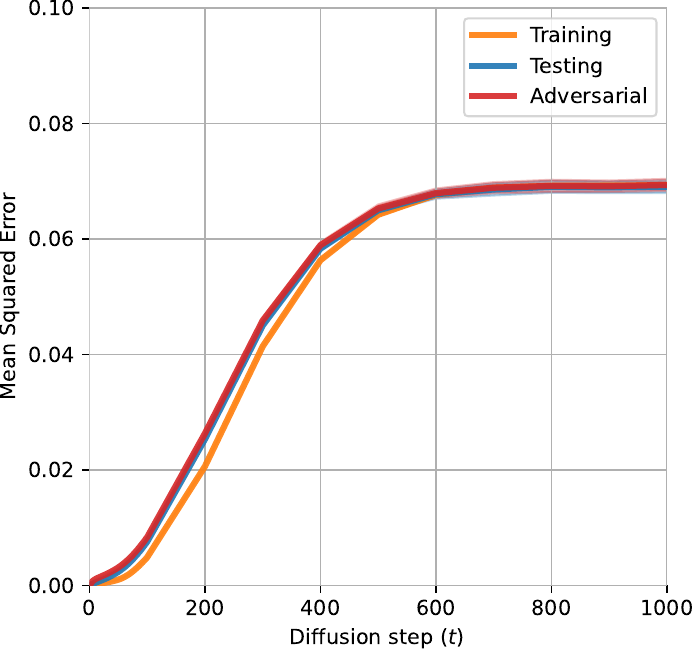}
  \caption{NSL-KDD}
  \label{fig:loss_t_NSL-KDD}
\end{subfigure}
\caption{Reconstruction loss over the diffusion steps $t$}
\label{fig:loss_t}
\end{figure}

%% Accuracy over T
\begin{figure}[t]
\centering
\begin{subfigure}{.5\textwidth}
  \centering
  \includegraphics[width=\linewidth]{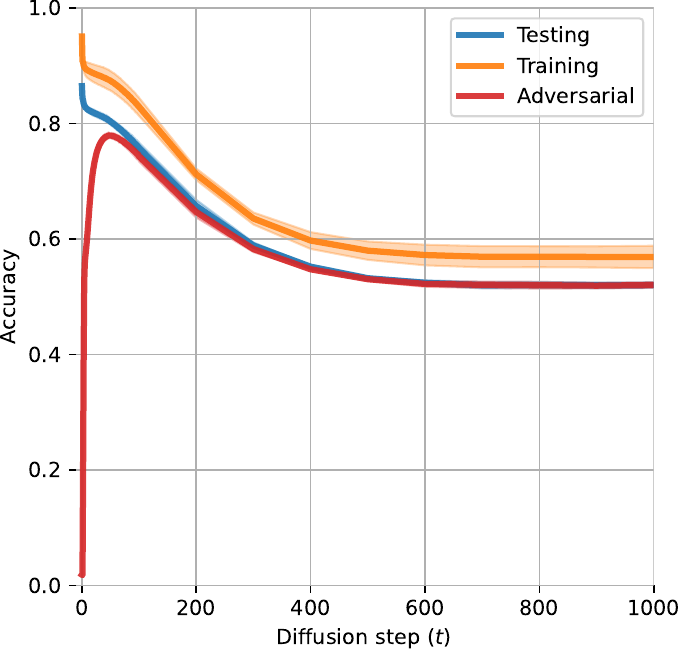}
  \caption{UNSW-NB15}
  \label{fig:acc_t_UNSW-NB15}
\end{subfigure}%
\begin{subfigure}{.5\textwidth}
  \centering
  \includegraphics[width=\linewidth]{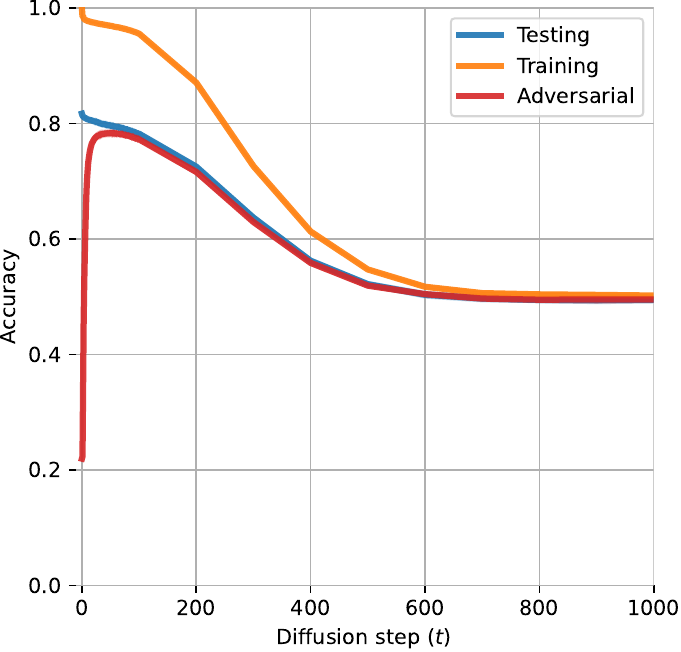}
  \caption{NSL-KDD}
  \label{fig:acc_t_NSL-KDD}
\end{subfigure}
\caption{Intrusion detection accuracy over the diffusion steps $t$}
\label{fig:acc_t}
\end{figure}

\subsubsection{Diffusion neural network size.}
Here, we analyze the training of diffusion models on network data.
We specifically compare the impact of the size of the diffusion neural network on the training loss.
We elaborate further on the impact of the number of diffusion steps on the training loss in Appendix \ref{appendix}.
Figure \ref{fig:loss_epochs_h} shows the reconstruction loss (in MSE) over the training epochs for different Units Per Layer (UPL).
%The blue line, green line, orange line, and red line represent a diffusion neural network with 10 hidden layers of 512, 960, 1024, and 2048 Units Per Layer (UPL) respectively.
%All diffusion models are trained with $\beta_{1}=10^{-4}$, $\beta_{T}=0.02$ and $T=1000$.
We observe that the reconstruction loss decreases earlier for larger diffusion neural networks.
Furthermore, larger neural networks reach lower MSE values when they stabilize, $0.008$ and $0.007$ with $2048$ UPL against $0.362$ and $0.026$ with $512$ UPL on UNSW-NB15 and NSL-KDD, respectively. 
This is due to the capacity of larger neural networks to model more complex patterns and learn better representations of the data.
However, larger models have drawbacks; they take significantly longer to train and require more computational power.
Larger models also display instabilities during the training; Figure \ref{fig:loss_epochs_h} shows how the 2048 UPL diffusion model loss peaks randomly between $150,000$ and $200,000$ epochs on UNSW-NB15.  
%% Give numbers about the training time
For the rest of the experiments, we will rely on diffusion neural networks with 10 hidden layers of 960 UPL trained for $200,000$ epochs.

\subsubsection{Reconstruction loss and accuracy over $t$.}

Figure \ref{fig:loss_t} shows the reconstruction loss over the diffusion steps $t$,
%The reconstruction loss consists of the MSE between $x_{0}$ and $\hat{x}_{0}$—its image after $t$ steps of the forward process and $t$ steps of the reverse process.
while Figure \ref{fig:acc_t} shows the accuracy of the intrusion detection model on the same reconstructed data over the diffusion steps $t$.
%The orange, blue, and red lines represent the training set, the testing set, and the adversarial examples generated from the testing set, respectively.
% The diffusion parameters used here are: $\beta_{1}=10^{-4}$, $\beta_{T}=0.02$, and $T=1000$.
%% , and a diffusion neural network with 10 hidden layers of 960 units.

Figure \ref{fig:loss_t} shows that the reconstruction loss increases similarly for all three sets, the lines even overlap on UNSW-NB15.
Indeed, as noise is added gradually, the data structure is slowly destroyed.
The more diffusion steps are applied, the more noise is added, and the harder it is to reconstruct precisely the data.
After $600$ steps, the reconstruction loss plateaued around $0.03$ and $0.07$ on UNSW-NB15 and NSL-KDD, respectively.
%% The reconstructed data are then equivalent to Gaussian noise with a mean $\mu=x_{0}$ and standard deviation $\sigma=\sqrt{\sum_{i=1}^{i=T} \beta_{i}^{2}}$, which indicates that all the data structure has been destroyed.
%% reconstruction based detection
Moreover, we notice that the reconstruction loss on adversarial examples is slightly superior to that on the original testing set.
The difference in the reconstruction losses can also be used as an indicator for detecting adversarial examples.
This avenue is not investigated in this paper, as adversarial detection approaches are out of our scope.

The accuracy curve in Figure \ref{fig:acc_t} corroborates the previous results: the training and testing accuracy decrease as the diffusion step increases due to the damaged data structure.
It becomes more challenging for the intrusion detection model to distinguish benign and malicious traffic, which decreases its accuracy.
After $600$ steps, the reconstructed data is too noisy to be classified correctly.
%%; the classification becomes random.  
%% both plateau after $600$ steps around $0.56$ and $0.52$ respectively.
%% Why these values?

\subsubsection{Purification performance.}
In order to evaluate the robustness of the intrusion detection after the diffusion model's purification, we focus on the red line in Figure \ref{fig:acc_t}, which represents the accuracy of the intrusion detection model on adversarial examples.
Those adversarial examples were generated from the testing set using the targeted Fast Gradient Sign Method (FGSM) with a perturbation amplitude $\epsilon=0.03$.
At step $0$, before purification, the accuracy on adversarial data is $0.02$, while it is $0.86$ on the original test data, indicating that the adversarial attack succeeded in misleading the intrusion detection model.
After a few diffusion steps, the adversarial accuracy increases drastically.
The added diffusion noise dilutes the adversarial perturbation that misleads the model while preserving enough data structure for a good classification.
After $44$ steps, the adversarial accuracy peaks at $78\%$ while the test accuracy is $80\%$. 
This diffusion step  $t^{*}$  is optimal as it maximizes the accuracy on adversarial examples while minimizing the impact on the non-adversarial test data.
This result empirically shows the purification capabilities of diffusion models in intrusion detection.

After the peak, the structure damage due to the addition of diffusion noise decreases the adversarial accuracy.
Since the adversarial perturbation has been removed, the difference between the test and adversarial accuracy disappears; it is below $0.01$ after $t=90$.
Both values decrease until they reach random classification when the reconstructed data is too noisy.

%% Beta_1 
\begin{figure}[t]
    \centering
    \begin{subfigure}{.5\textwidth}
      \centering
      \includegraphics[width=\linewidth]{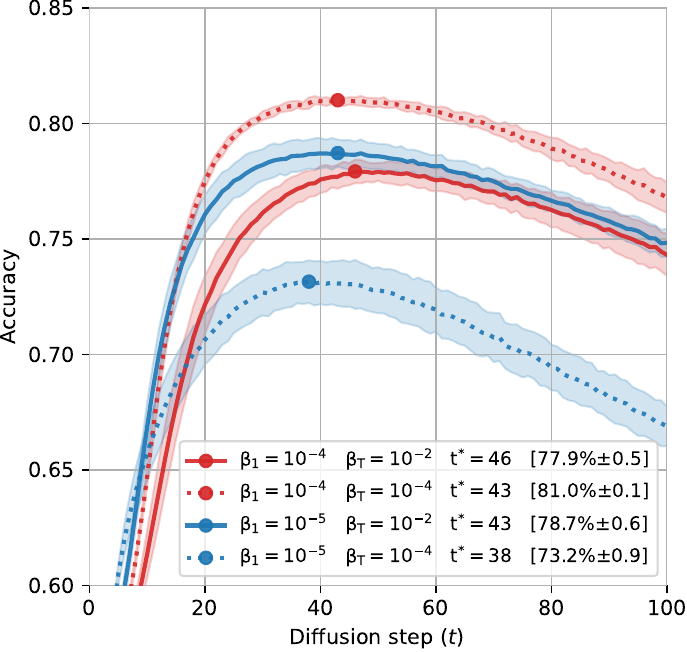}
      \caption{UNSW-NB15}
      \label{fig:acc_t_B1_UNSW-NB15}
    \end{subfigure}%
    \begin{subfigure}{.5\textwidth}
      \centering
      \includegraphics[width=\linewidth]{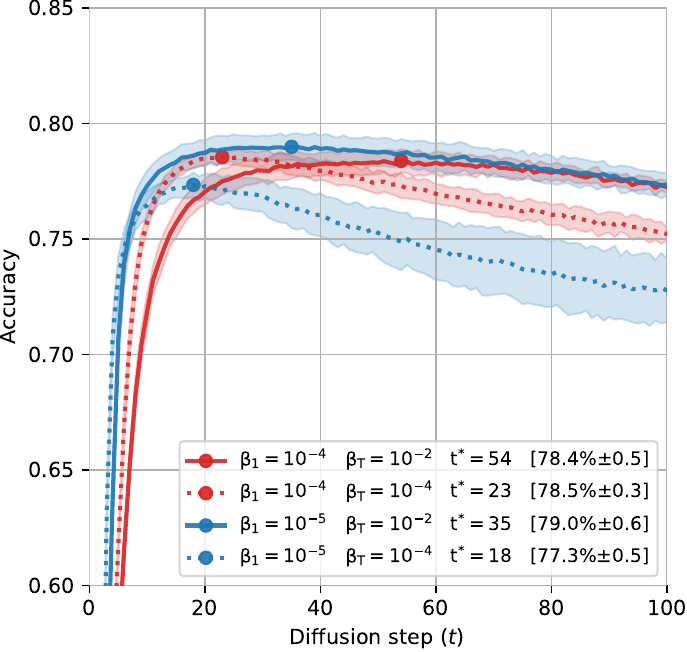}
      \caption{NSL-KDD}
      \label{fig:acc_t_B1_NSL-KDD}
    \end{subfigure}
    \caption{Intrusion detection accuracy over diffusion step $t$ for different $\beta_{1}$. Continuous lines for $\beta_{T}=10^{-2}$ and dotted lines for $\beta_{T}=10^{-4}$. The marker indicates the maximum accuracy reached at the optimal diffusion step $t^{*}$.}
    \label{fig:acc_t_B1}
\end{figure}

\subsubsection{Variance schedule.}
The variance of the Gaussian noise added at step $t$ of the diffusion process is denoted $\beta_{t}$.
It follows a linear distribution of $T$ evenly spaced values between $\beta_{1}$ and $\beta_{T}$.
The variance schedule is an essential parameter of the diffusion process; we hypothesize that it is also critical to the purification capabilities of diffusion models.
In the following, we study the impact of the variance schedule $\beta$ on the purification capabilities of the diffusion model through the accuracy of the intrusion detection model on purified examples.
Using a fixed $T=1000$, we vary both $\beta_{1}$ and $\beta_{T}$ to find an optimal schedule.

Figure \ref{fig:acc_t_B1} shows the impact of $\beta_1$, the first value of the variance schedule.
%The red lines and blue lines correspond to $\beta_1 = 10^{-4}$ and $\beta_1 = 10^{-5}$ respectively.
If we focus on the continuous lines, corresponding to $\beta_{T}=10^{-2}$, we do not see a significant difference between the two values of $\beta_{1}$.
However, with a smaller $\beta_{T}$, the difference between $\beta_{1}$ becomes significant.
The dotted lines, corresponding to $\beta_{T} = 10^{-4}$, show a large difference between the two values of $\beta_{1}$.
As the final $\beta$ decreases, the difference between the accuracy of the two $\beta_{1}$ values increases.
The impact of the initial variance $\beta_{1}$ is therefore linked to the length of the variance schedule $\beta_{T} - \beta_{1}$ and becomes less significant as the interval increases. 
This result suggests that $\beta_{T}$ plays an influential role in the purification.

%% Beta_T
\begin{figure}[t]
    \centering
    \begin{subfigure}{.5\textwidth}
      \centering
      \includegraphics[width=\linewidth]{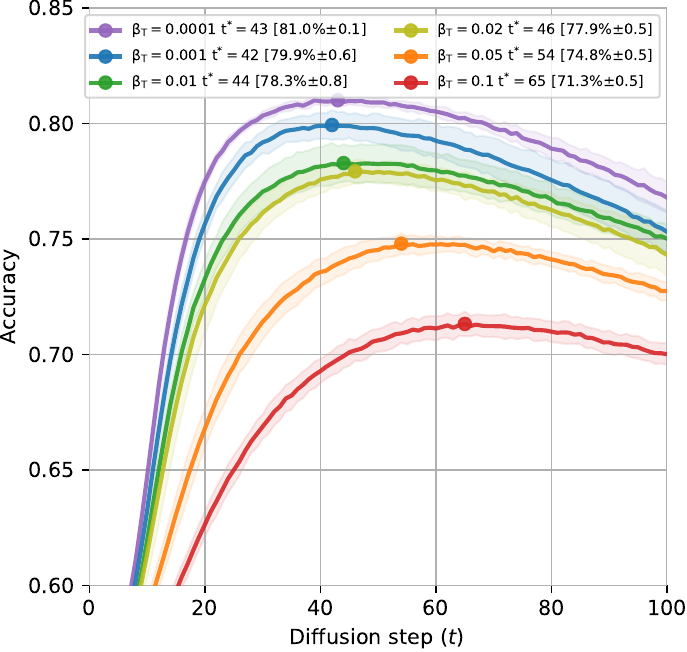}
      \caption{UNSW-NB15}
      \label{fig:acc_t_BT_UNSW-NB15}
    \end{subfigure}%
    \begin{subfigure}{.5\textwidth}
      \centering
      \includegraphics[width=\linewidth]{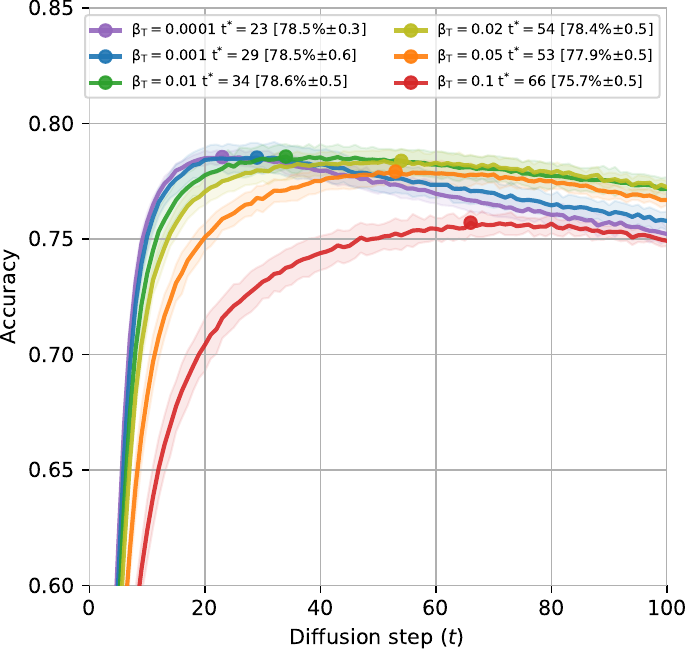}
      \caption{NSL-KDD}
      \label{fig:acc_t_BT_NSL-KDD}
    \end{subfigure}
    \caption{Intrusion detection accuracy over diffusion step $t$ for different $\beta_{T}$}
    \label{fig:acc_t_BT}
\end{figure}

Figure \ref{fig:acc_t_BT} shows the impact of $\beta_{T}$, the final variance of the diffusion schedule.
%Each line represents the accuracy of the intrusion detection model on purified adversarial examples from the testing set, the marker indicates the maximum accuracy reached at the optimal diffusion step $t^{*}$.
The figure shows an increasing accuracy with smaller $\beta_{T}$ values.
The intuition is that as $\beta_{T}$ decreases, the interval between successive variance values decreases, which makes the noising more gradual and easier for the neural network to reconstruct.
On UNSW-NB15, the maximum accuracy value is $81\%\pm0.1$ at $t^{*}=43$, it was reached with the smallest value $\beta_{T}=10^{-4}$, which represents a constant variance schedule (since $\beta_{1}=10^{-4}$).
On NSL-KDD, the maximum accuracy value is very close between when $\beta_{1}\leq10^{-2}$, but $\beta_{1}\leq10^{-4}$ is the earliest to achieve $78.5\pm0.3$ after only $t^{*}=23$.  
%The NSL-KDD example shows well how a smaller $\beta_{T}$ helps to reach an optimum with fewer diffusion steps.

\subsubsection{Number of diffusion steps $T$.}
In addition to the initial and final perturbation values $\beta_{1}$ and $\beta_{T}$, the diffusion process is characterized by the number of diffusion steps $T$.
This parameter determines the granularity of the diffusion since a larger number of steps $T$ makes the step size smaller.
%% , as shown in Equation \ref{eq:diffusion_step}:
%% \begin{equation}
%%     \beta_{t} - \beta_{t-1} = \frac{\beta_{T} - \beta_{1}}{T}
%% \label{eq:diffusion_step}
%% \end{equation}

In the following, we study how the number of diffusion steps $T$ affects the optimal diffusion steps $t^{*}$.
We compare diffusion models with $T=100$ and $T=1000$ with respect to the optimal diffusion step $t^{*}$ and identify how it translates to an equivalent amount of noise.
In Appendix \ref{appendix}, we further investigate the impact of the number of diffusion steps $T$ on the training loss and the purification performance with the optimal variance schedule.

%% t B_t V_t
\begin{figure}[t]
    \centering
    \begin{subfigure}[t]{.33\textwidth}
      \centering
      \includegraphics[width=\linewidth]{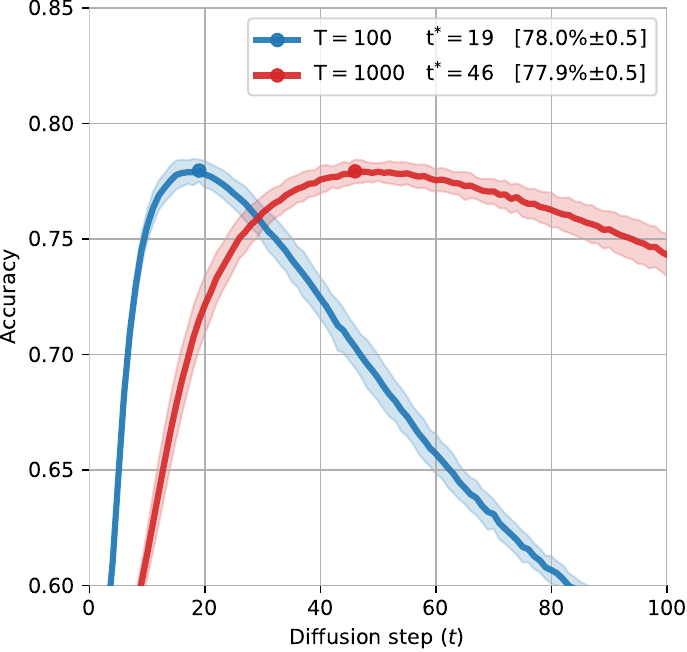}
      \label{fig:acc_t_T_UNSW-NB15}
    \end{subfigure}%
    \begin{subfigure}[t]{.33\textwidth}
      \centering
      \includegraphics[width=\linewidth]{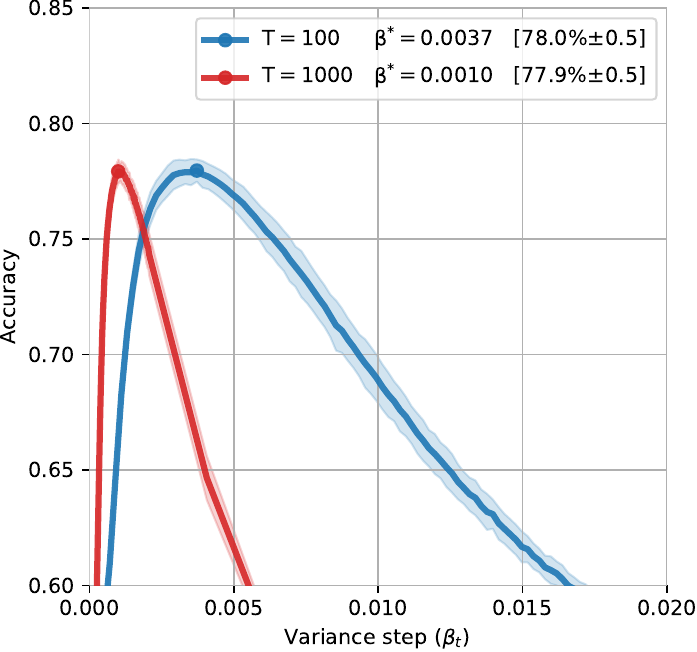}
      \label{fig:acc_B_T_UNSW-NB15}
      \caption{UNSW-NB15}
    \end{subfigure}
    \begin{subfigure}[t]{.33\textwidth}
      \centering
      \includegraphics[width=\linewidth]{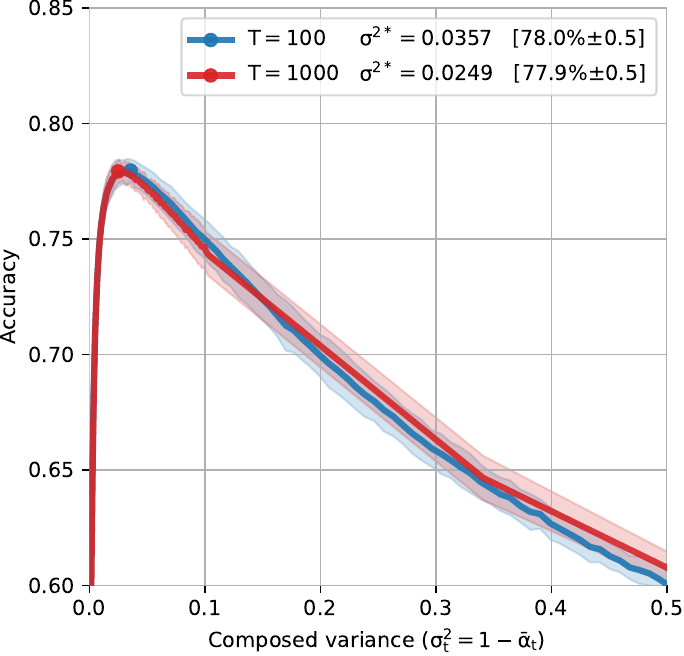}
      \label{fig:acc_V_T_UNSW-NB15}
    \end{subfigure}
\medskip
    \centering
    \begin{subfigure}[t]{.33\textwidth}
      \centering
      \includegraphics[width=\linewidth]{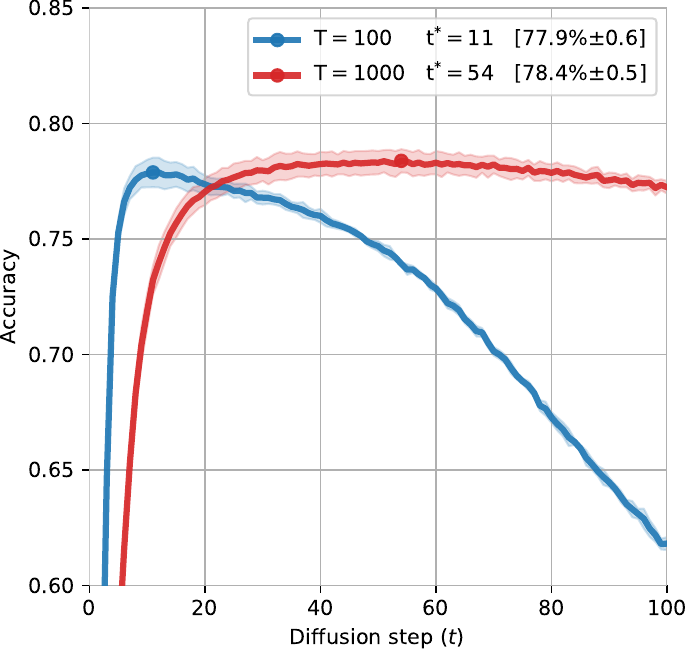}
      \label{fig:acc_t_T_NSL-KDD}
    \end{subfigure}%
    \begin{subfigure}[t]{.33\textwidth}
      \centering
      \includegraphics[width=\linewidth]{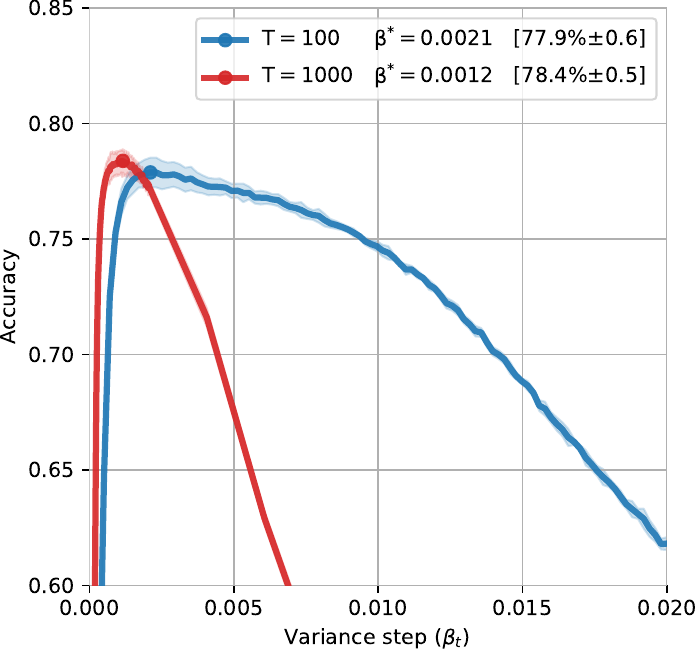}
      \label{fig:acc_B_T_NSL-KDD}
      \caption{NSL-KDD}
    \end{subfigure}
    \begin{subfigure}[t]{.33\textwidth}
      \centering
      \includegraphics[width=\linewidth]{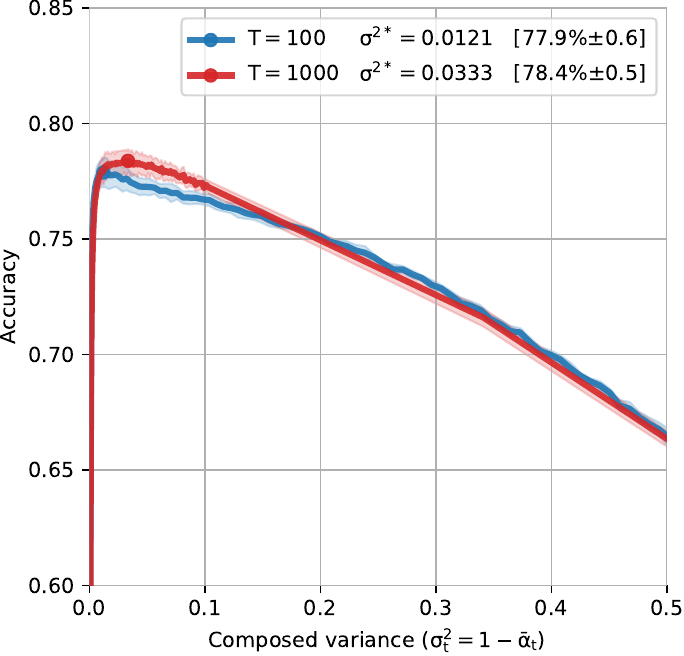}
      \label{fig:acc_V_T_NSL-KDD}
    \end{subfigure}
    \caption{Intrusion detection accuracy over $t$, $\beta_t$, and $\sigma^{2}_t$ for different number of diffusion steps $T$}
    \label{fig:acc_T}  
\end{figure}

Figure \ref{fig:acc_T} shows the impact of the number of diffusion steps $T$ over three references: the diffusion step $t$, the variance step $\beta_t$, and the composed variance $\sigma^{2}_{t}$.
%Each line represents the accuracy of the intrusion detection model on purified adversarial examples from the testing set, the marker indicates the maximum accuracy.
%Here we use the standard variance interval $\beta_{1}=10^{-4}$ and $\beta_{T}=0.02$, for which the performance of $T=1000$ and $T=100$ are similar.
This experiment uses the standard diffusion parameters to focus the analysis on when the optimum is recorded rather than its value.
With the optimal variance interval recorded in Figure \ref{fig:acc_t_BT}, $\beta_{1}=\beta_{T}=10^{-4}$, the 1000-step diffusion model largely over-performs the 100-step one on UNSW-NB15, as shown in Figure \ref{fig:acc_t_T_BT} (Appendix \ref{appendix}).

In the first part of Figure \ref{fig:acc_T}, we see the accuracy across the diffusion step $t$.
The $T=100$ diffusion model reaches its optimum at $t^{*}=19$ and $t^{*}=11$, while the $T=1000$ reaches its optimum at $t^{*}=46$ and $t^{*}=54$ on UNSW-NB15 and NSL-KDD, respectively.
This indicates a slower evolution when $T=1000$, which is coherent since the variance steps are smaller.
Therefore, we plot the same values with respect to the variance steps $\beta_{t}$ to find a similarity.
In the second part of Figure \ref{fig:acc_T}, the x-axis corresponds to $\beta_{t}$ and covers the whole interval from $0$ to $\beta_{T}=0.02$.
However, the optimum is still reached at distant values of $\beta$.
For $T=100$, it is reached with $\beta^{*}=0.0037$ and $\beta^{*}=0.0021$, while for $T=1000$, it is reached with $\beta^{*}=0.0010$ and $\beta^{*}=0.0012$ in UNSW-NB15 and NSL-KDD, respectively. 

$\beta_{t}$ is the amount of variance applied at the diffusion step $t$, but it does not correspond to the total variance applied to the initial data $x_0$.
Indeed, the diffusion forward process is a composition of Gaussian distributions with gradual variances $\beta_{t}$.
The total (composed) variance $\sigma^{2}_{t}$ of such composition is the sum of the individual variances, $\sigma^{2}_{t} = 1 - \bar{\alpha}_t$ (Equation \ref{eq:transition_cumulative_forward}).
%\begin{equation*}
%    q\left(x_{t} | x_0 \right) \simeq \mathcal{N}\left( x_{0} ; \sqrt{\bar{\alpha_{t}}} x_{0}, \left(1 - \bar{\alpha}_t\right) I \right)
%\end{equation*}

%\begin{equation*}
%    \bar{\alpha} = \prod_{i=1}^{i=t}{1 - \beta_{t}} 
%\end{equation*}

%\begin{equation*}
%    \sigma^{2}_{t} = \sqrt{1 - \bar{\alpha}_t}
    %%\label{eq:composed_variance}
%\end{equation*}

The third part of Figure \ref{fig:acc_T} shows the accuracy across the composed variance $\sigma_t$ where the two lines overlap, indicating a similar accuracy regardless of the number of diffusion steps $T$.
The optimum is reached at $\sigma^{2}_{t}=0.0249$ and $\sigma^{2}_{t}=0.0333$ for $T=1000$, and $\sigma^{2}_{t}=0.0357$ and $\sigma^{2}_{t}=0.0121$ for $T=100$ on UNSW-NB15 and NSL-KDD, respectively.
Considering the scale of $\sigma^{2}_{T}$ in Figure \ref{fig:acc_T}, the optimum values are relatively close.
We note from this experiment that the optimal noise added ${\sigma^{2}}^{*}$ approaches $0.03$, which corresponds to the adversarial perturbation amplitude $\epsilon$ used in these experiments.
This result suggests a dependence between the optimal noise amount and the perturbation amplitude.

%% Epsilon
\begin{figure}[t]
\centering
\begin{subfigure}{.5\textwidth}
  \centering
  \includegraphics[width=\linewidth]{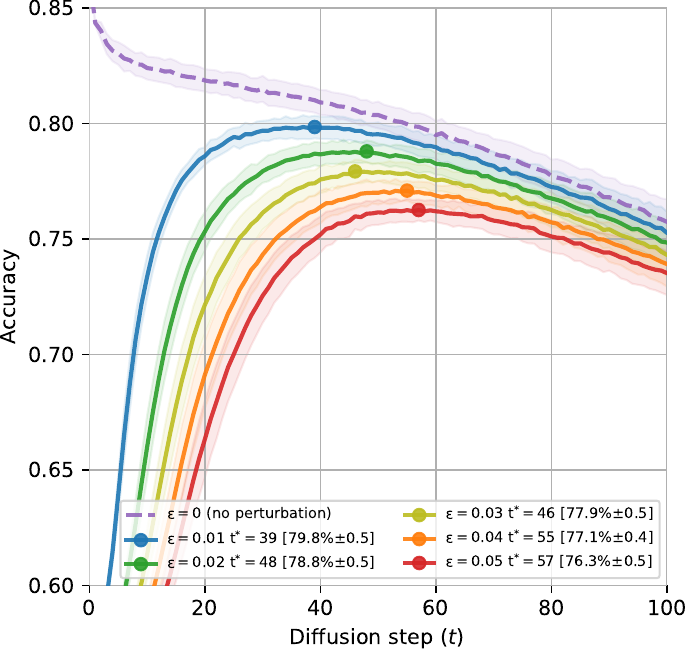}
  \caption{UNSW-NB15}
  \label{fig:acc_t_epsilon_UNSW-NB15}
\end{subfigure}%
\begin{subfigure}{.5\textwidth}
  \centering
  \includegraphics[width=\linewidth]{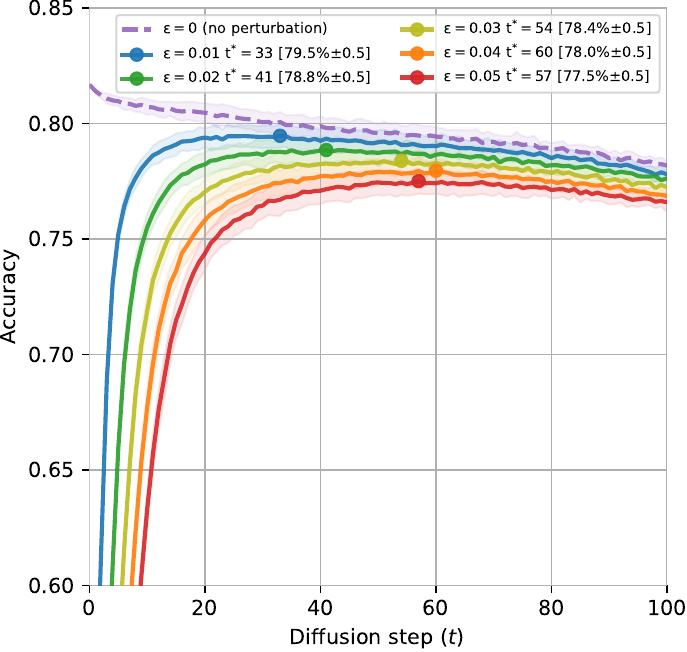}
  \caption{NSL-KDD}
  \label{fig:acc_t_epsilon_NSL-KDD}
\end{subfigure}
\caption{Intrusion detection accuracy over the diffusion step $t$ for different adversarial perturbation amplitudes $\epsilon$ generated with FGSM}
\label{fig:acc_t_epsilon}
\end{figure}

\subsubsection{Adversarial perturbation amplitude $\epsilon$.}

Beyond the diffusion parameters, the purification performance of diffusion models depends on the amount of adversarial perturbation $\epsilon$ added to the data.
Figure \ref{fig:acc_t_epsilon} shows the accuracy of the intrusion detection model on increasing $\epsilon$ values.
%% for both UNSW-NB15 and NSL-KDD.
%% We use the standard diffusion parameters $T=1000$, $B_{1}=10^{-4}$, and $B_{T}=0.02$.
%% he adversarial examples are generated with targeted FGSM and the perturbation amplitude ranges from $0$ (clean test data) to $0.05$
It demonstrates how the purification performance decreases as the perturbation amplitude increases.
The accuracy at $t^{*}$ goes from $79.8\%\pm0.5$ and $79.5\%\pm0.5$ when $\epsilon=0.01$ to $76.3\%\pm0.5$ and $77.5\%\pm0.5$ when $\epsilon=0.05$ on UNSW-NB15 and NSL-KDD, respectively.
Another pattern is that it takes more diffusion steps to reach an optimum as $\epsilon$ increases.
The two phenomena are linked: the diffusion model needs to add more perturbation to dilute a larger $\epsilon$, thus taking more diffusion steps.
However, the test accuracy (purple line) decreases as more noise is added.
Since it represents an upper bound on the adversarial accuracy, it causes the optimal adversarial accuracy to decrease as $\epsilon$ increases.
%% More impact on UNSW-NB15 than on NSL-KDD

\subsubsection{Adversarial attacks.}
The efficiency of adversarial examples also depends on the method used to generate them.
Various methods exist, each optimizing different distance norms and criteria.
While previous experiments show how diffusion models considerably improve the adversarial accuracy of intrusion detection models, they have only been tested on one adversarial attack (FGSM).
Here, we study how the purification performance is generalized to other adversarial examples' generation methods.
We compare five well-established methods, namely: DeepFool~\cite{moosavi_dezfooli_deepfool_2016}, Jacobian-based Saliency Map Attack (JSMA)~\cite{papernot_limitations_2016}, Fast Gradient Sign Method (FGSM)~\cite{goodfellow_explaining_2015}, Basic Iterative Method (BIM)~\cite{kurakin_adversarial_2017}, and Carlini\&Wagner's $L_{2}$ attack~\cite{carlini_towards_2017}.

%% Different attacks
\begin{figure}[t]
\centering
\begin{subfigure}{.5\textwidth}
  \centering
  \includegraphics[width=\linewidth]{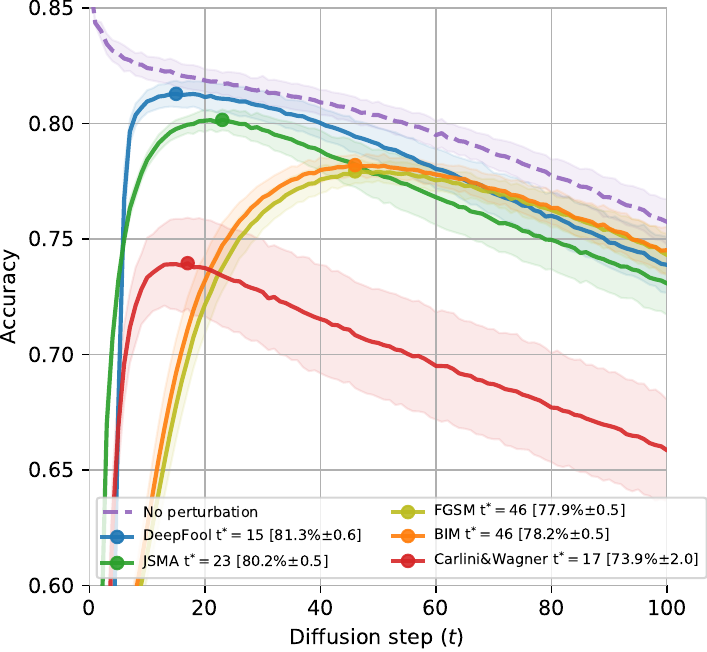}
  \caption{UNSW-NB15}
  \label{fig:acc_t_attack_UNSW-NB15}
\end{subfigure}%
\begin{subfigure}{.5\textwidth}
  \centering
  \includegraphics[width=\linewidth]{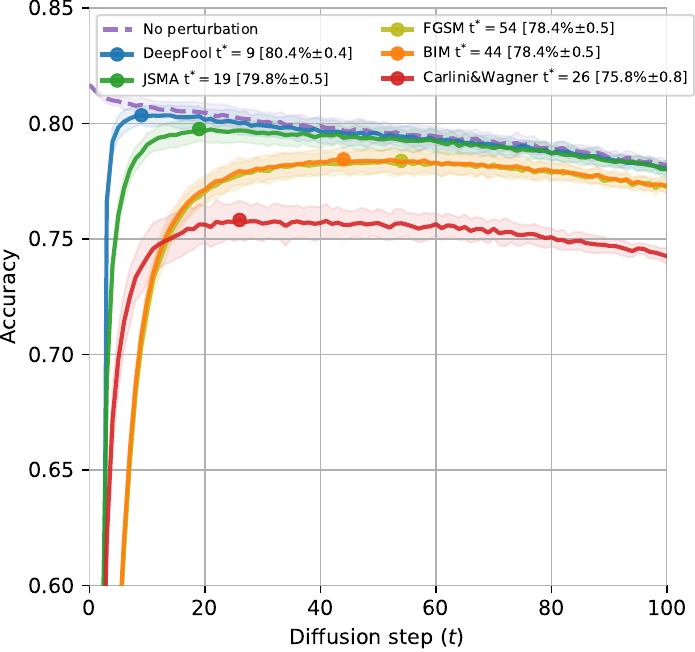}
  \caption{NSL-KDD}
  \label{fig:acc_t_attack_NSL-KDD}
\end{subfigure}
\caption{Accuracy over the diffusion step $t$ for different adversarial attacks}
\label{fig:acc_t_attack}
\end{figure}

Figure \ref{fig:acc_t_attack} shows the adversarial accuracy of the intrusion detection model for the adversarial attacks and the testing set baseline.
%% The diffusion parameters used here are the standard $T=1000$, $B_{1}=10^{-4}$, and $B_{T}=0.02$. 
We note that the diffusion models successfully purify adversarial examples from all 5 generation methods.
However, the optimal adversarial accuracy and diffusion step vary from one method to the other.
Carlini\&Wagner's attack is the most resistant to adversarial purification, with a maximum adversarial accuracy of $73.9\%\pm2$ and $75.8\%\pm0.8$ on UNSW-NB15 and NSL-KDD, respectively.
The optimum is reached after only $17$ steps on UNSW-NB15, $2.7$ times faster than FGSM.
On NSL-KDD, the adversarial accuracy is stable between $30$ and $100$, making $t^{*}$ less accurate.
%% small t = higher test accuracy

FGSM and BIM achieve similar performance; FGSM reaches a maximum of $77.9\%\pm0.5$ and $78.4\%\pm0.5$, while BIM reaches a maximum of $78.2\%\pm0.5$ and $78.4\%\pm0.5$ on UNSW-NB15 and NSL-KDD, respectively.
The optimal diffusion steps are also very close between the two.
This similarity is explained by the fact that BIM is based on FGSM; it applies small FGSM steps iteratively (100 in our experiments) to optimize the adversarial examples.
However, the iterative approach does not make BIM's adversarial examples more robust to our diffusion-based adversarial purification.
The diffusion models achieve the highest purification performance on DeepFool's adversarial examples, with a maximum accuracy of $81.3\%\pm0.6$ and $80.4\%\pm0.4$ on UNSW-NB15 and NSL-KDD, respectively, closely followed by JSMA with a maximum accuracy of $80.2\%\pm0.5$ and $79.8\%\pm0.5$ on UNSW-NB15 and NSL-KDD, respectively.
%% Make sure it is the fastest
DeepFool and JSMA also had the fastest-growing adversarial accuracy values among other attacks, almost reaching the testing accuracy upper bound.

\section{Discussion}
\label{section: discussion}
\paragraph{Diffusion neural network.}
The size of the neural network has a considerable impact on the reconstruction loss of the diffusion model.
Larger neural networks converge faster and reach lower reconstruction loss values due to their capacity to model more complex patterns and learn better representations.
The main obstacle to using larger neural networks is that they take longer to process the data, making the training and reconstruction longer.
%details about the time
The diffusion neural network should be hyperparameterized with particular attention to domain-specific time constraints, especially in network intrusion detection, where the reaction time is critical.

In addition to the size of the diffusion neural network, other hyperparameters impact the reconstruction loss and, potentially, the purification performance.
%Increasing the depth of the neural network helps capture more complex relationships between the features.
The choice of the loss function, the optimization algorithm, and the learning rate affect the convergence time and help find better optimums.
Finally, fully connected neural networks can be replaced with more sophisticated neural network architectures that model the diffusion process more precisely.
In sum, optimizing diffusion neural networks is a promising avenue for improving the reconstruction loss, the purification performance, and even the processing time of diffusion models.

\paragraph{Variance schedule $\beta$.}
The variance schedule determines the amount of noise added at each step of the forward diffusion process.
In the experiments, the schedule is linear—it consists of $T$ evenly space values between $\beta_{1}$ and $\beta_{T}$.
This schedule is essential for adversarial purification, since adding too little noise does not remove the adversarial perturbation, and too much noise corrupts the data structure.
The importance of $\beta_{1}$, the variance of the first diffusion step, depends on the total length of the diffusion schedule $\beta_{T} - \beta_{1}$.
When the schedule is small, the choice of the starting value makes a considerable difference in the purification capabilities.
As the difference between $\beta_{1}$ and $\beta_{T}$ increases, $\beta_{1}$ is less significant since it has little impact on the rest of the variance values.

On the other hand, the last value $\beta_{T}$ has a more significant impact on the variance schedule, which also extends to the purification performance.
In the standard setting, where $T=1000$,  $\beta_{1}=10^{-4}$, and $\beta_{T}=0.02$, a smaller value of $\beta_{T}$ leads to a better adversarial accuracy.
The best purification performance was recorded with the smallest value $\beta_{T}=\beta_{1}=10^{-4}$,  which describes a constant variance schedule.
%% However, These results depend on the settings of the experiments, they may not extend to other settings.

As opposed to generative models, diffusion models in adversarial purification do not require the forward process to reach a Gaussian distribution $\mathcal{N}\left( x_{T} ; 0, I_n\right)$  after $T$ steps.
Therefore, there is more freedom in choosing the diffusion parameters to optimize the adversarial accuracy.

\paragraph{Number of diffusion steps $T$.}
The number of diffusion steps affects multiple aspects of the diffusion process.
It determines the size of the variance schedule $\beta$ and its values since there are $T$ evenly spaced values.
In the best-performing case $\beta_{T}=\beta_{1}=10^{-4}$, the number of diffusion steps makes a considerable difference in favor of a larger $T$.
Having the same variance schedule divided into more variance steps allows the neural network to learn more details about the applied noise, which are then helpful to accurately reconstruct the data in the reverse process.
%% This justifies the lower reconstruction loss in Figure \ref{fig:loss_epochs_t}.

The number of diffusion steps also affects the optimal diffusion step $t^{*}$; it happens earlier in the process when $T$ is smaller.
This is explained by the fact that the diffusion steps $t$ do not correspond to an equivalent variance step $\beta_t$.
However, even when considering $\beta_t$ instead of $t$, the reconstruction curve does not match.
This is because the $\beta_t$ values are gradually added to the data (Equation \ref{eq:transition_one_step_forward}) according to the variance step; they do not represent the total variance applied to the data.
Instead, we consider the variance of the composition, $\sigma_{t}^{2} = 1 - \bar{\alpha}$, which makes the values of $T=100$ and $T=1000$ match.
The total variance that maximizes the adversarial accuracy, ${\sigma^{2}}^{*}$ is very close between the two different $T$ values.
Furthermore, the optimal variance ${\sigma^{2}}^{*}$ approaches the value of the perturbation amplitude $\epsilon$, demonstrating the dependence of the diffusion noise on the adversarial perturbation.
%On the other side, this quantity largely depends on the nature of the adversarial examples—the amount of perturbation $\epsilon$ and the adversarial generation method.

In sum, diffusion models with more diffusion steps achieve better adversarial accuracy with an optimal variance schedule.
However, it takes more steps to reach optimum, translating into a longer purification time.
Thus, the choice of $T$ should also consider the time constraints that characterize the application domain.

\paragraph{Adversarial perturbation amplitude.}
The diffusion-based adversarial purification effectively removes adversarial perturbations from network traffic data.
However, the effectiveness of this method varies slightly depending on the nature of the adversarial examples.
In the case of adversarial examples generated with FGSM, the parameter $\epsilon$ determines the amount of perturbation added to the data.
%% $L_{\inf}$ norm
As the $\epsilon$ increases, more noise is required to dilute the perturbation, making the optimum occur later in terms of diffusion steps.
Another side effect is that the extra required noise also dilutes some of the data structure, thus decreasing the test accuracy, representing an upper bound to the adversarial accuracy.
Therefore, the maximum adversarial accuracy decreases as $\epsilon$ increases.

\paragraph{Adversarial attacks.}
In addition to the perturbation amplitude $\epsilon$, the purification process is sensitive to the adversarial examples' generation method.
The results show a considerable gap in the adversarial accuracy between different methods.
In particular, DeepFool and JSMA are easier to purify and approach the test accuracy upper-bound after a few diffusion steps, FGSM and BIM achieve close performance due to their similarities, and Carlini\&Wagner's $L_{2}$ attack is the most resistant but our method still recovers up to $75\%$ of the accuracy.
%while still losing $87\%$ of its effectiveness after purification.

\paragraph{Adversary's constraints.}
The adversarial examples' generation method and the adversarial perturbation amplitude are both parameters controlled by the adversary when they generate the adversarial examples.
However, they are constrained in the amount of perturbation they can add.
A larger adversarial perturbation can increase the chances of being detected, cancel the purpose of the data, or even break its consistency, especially for highly structured data like network traffic~\cite{merzouk_investigating_2022}.
In the experiments, we consider the worst-case scenario where the attacker perturbs all the data features.
However, the design of real-world diffusion-based adversarial purification models should consider a realistic threat model to optimize the diffusion parameters.

\section{Conclusion}
\label{section: conclusion}

Diffusion models are a promising approach to adversarial purification.
Their seamless integration with existing systems and generalization across attack methods make them particularly interesting in the context of intrusion detection.

Throughout this paper, we have demonstrated the effectiveness of diffusion models in mitigating the threat of adversarial examples against intrusion detection.
We have compared several diffusion neural network sizes, which show that larger neural networks yield lower loss values despite their increased demands in time and computational resources.
Our analysis of the variance schedule indicates the importance of the final variance $\beta_{T}$ in determining the purification performance, with smaller values achieving the highest accuracy.
Furthermore, we have shown that the optimal amount of diffusion noise ${\sigma^{2}}^{*}$ is nearly constant regardless of the number of diffusion steps $T$ and that it approaches the value of the perturbation amplitude $\epsilon$. 
However, in terms of purification performance, diffusion models with a larger $T$ display better adversarial accuracy despite requiring more diffusion steps.
Finally, we benchmarked our method against five state-of-the-art adversarial attacks and an increasing perturbation amplitude.

While scalability and computational complexity remain the main challenges for diffusion models in intrusion detection, especially for inline detection, we envision future research endeavors to refine and optimize diffusion models for practical deployment.
As novel adversarial attacks emerge and challenge adversarial defenses~\cite{kang_diffattack_2023}, our future work will focus on adapting diffusion-based purification to these attacks.
Ultimately, complementing diffusion models with other defensive techniques remains necessary to prevent a single point of failure.

\section*{Acknowledgments}
This work was supported by \href{https://www.mitacs.ca/}{Mitacs} through the Mitacs Accelerate International program and the \href{https://www.critical.polymtl.ca/}{CRITiCAL} chair.
It was enabled in part by support provided by \href{https://calculquebec.ca/}{Calcul Québec}, \href{https://computeontario.ca/}{Compute Ontario}, the BC DRI Group, and the \href{https://alliancecan.ca/}{Digital Research Alliance of Canada}.

\bibliographystyle{splncs04}
\bibliography{main}

\appendix

\section{More on the number of diffusion steps $T$}
\label{appendix}

\begin{figure}[t]
\centering
\begin{subfigure}{.5\textwidth}
  \centering
  \includegraphics[width=\linewidth]{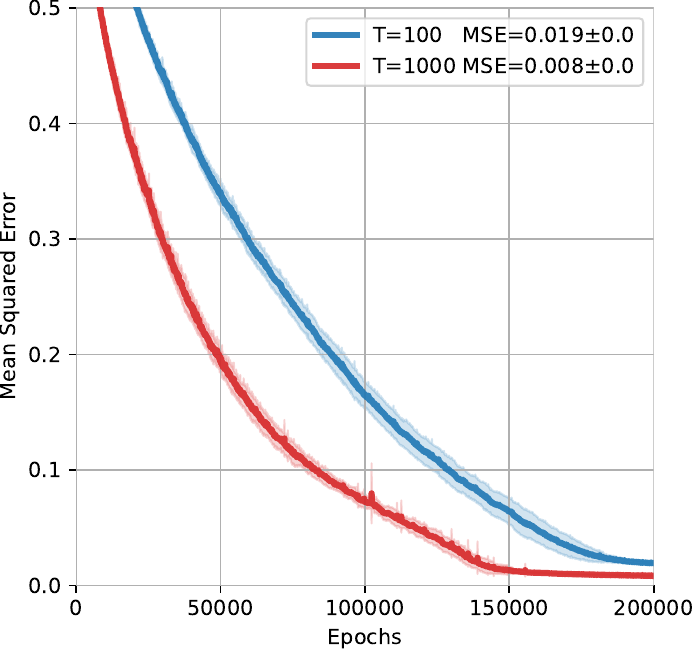}
  \caption{UNSW-NB15}
  \label{fig:loss_epochs_t_UNSW-NB15}
\end{subfigure}%
\begin{subfigure}{.5\textwidth}
  \centering
  \includegraphics[width=\linewidth]{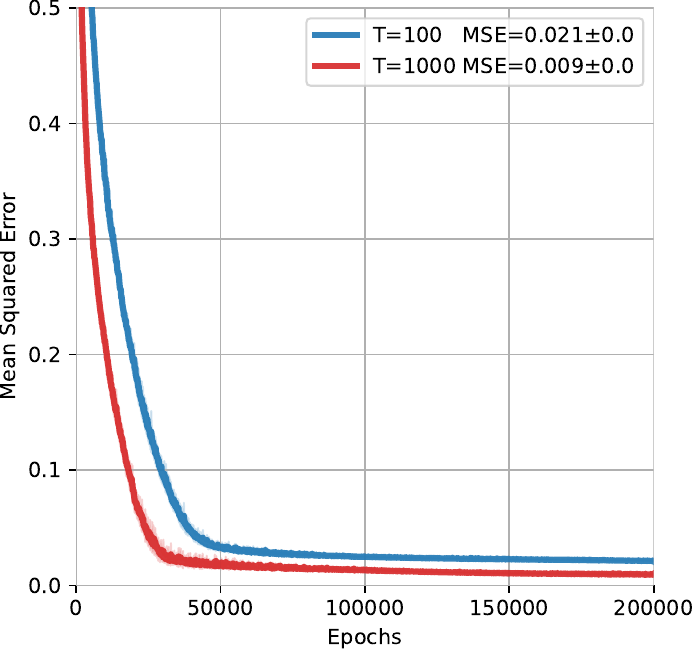}
  \caption{NSL-KDD}
  \label{fig:loss_epochs_t_NSL-KDD}
\end{subfigure}
\caption{Reconstruction loss over training epochs for $T=100$ and $T=1000$}
\label{fig:loss_epochs_t}
\end{figure}

Figure \ref{fig:loss_epochs_t} shows the reconstruction loss over the training epochs.
% The blue line and the red line represent a diffusion model with 100 and 1000 diffusion steps $T$, respectively.
Both diffusion models are trained with $\beta_{1}=10^{-4}$, $\beta_{T}=10^{-2}$, and a diffusion neural network with 10 hidden layers of 960 ReLU units. 
We observe that the reconstruction loss for $T=1000$ is always lower than for $T=100$.
Also, the loss curve for $T=1000$ stabilizes earlier (around $150,000$ epochs as opposed to $190,000$ epochs when $T=100$ on UNSW-NB15).
Despite both models having the same $\beta_{1}$ and $\beta_{T}$, a larger $T$ divides the range into smaller steps, allowing for a more gradual noising.
As the added noise increases more slowly, the diffusion neural network reconstructs better $x_{t-1}$ at each step.

%% Training cuvres for T

\begin{figure}[h]
\centering
\begin{subfigure}{.5\textwidth}
  \centering
  \includegraphics[width=\linewidth]{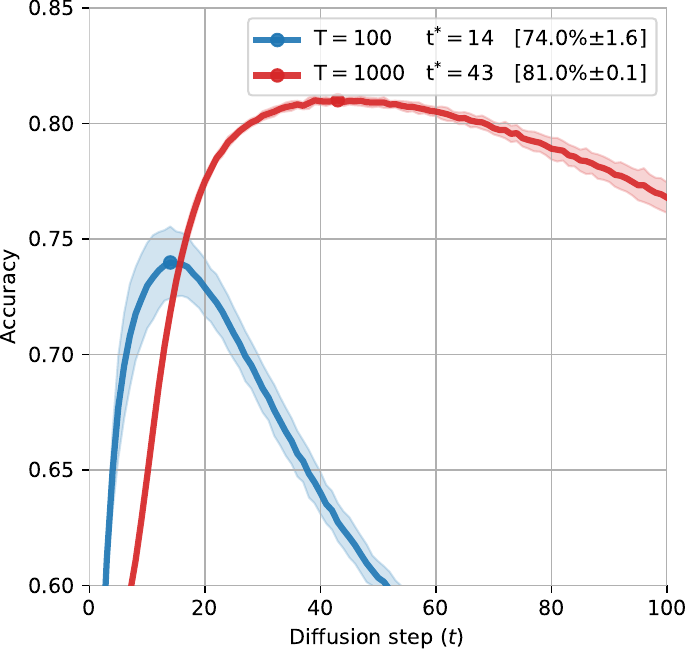}
  \caption{UNSW-NB15}
  \label{fig:acc_t_T_BT_UNSW-NB15}
\end{subfigure}%
\begin{subfigure}{.5\textwidth}
  \centering
  \includegraphics[width=\linewidth]{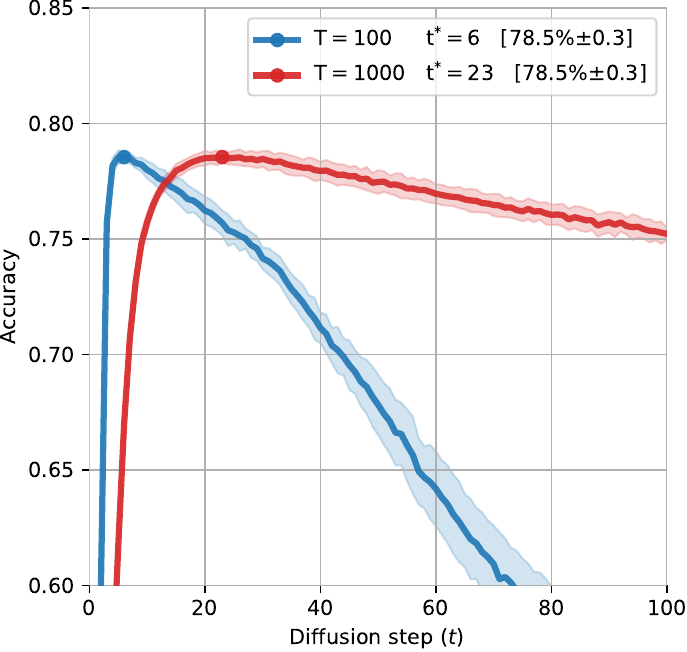}
  \caption{NSL-KDD}
  \label{fig:acc_t_T_BT_NSL-KDD}
\end{subfigure}
\caption{Accuracy over the diffusion step $t$ for different number of diffusion steps $t$ using the optimal variance interval recorded in Figure \ref{fig:acc_t_BT}: $\beta_{1}=\beta_{T}=10^{-4}$}
\label{fig:acc_t_T_BT}
\end{figure}

\end{document}